\documentclass[twocolumn]{aastex61}

\received{\today}
\revised{XXX}
\accepted{XXX}
\submitjournal{ApJ}

\shorttitle{influence of orbital resonances on the water transport in binary star systems}
\shortauthors{Bancelin et al.}

\usepackage{graphicx}	
\usepackage{amsmath}	
\usepackage{amssymb}	
\usepackage{slashbox}
\usepackage{color}
\usepackage{mathptmx} 
\usepackage{multirow}
\usepackage{wasysym}
\usepackage{url}
\usepackage{txfonts}
\def\apjl{ApJ}
\def\apjs{ApJS}
\def\ao{Appl.~Opt.}
\def\aap{A\&A}

\begin{document}

\title{The influence of orbital resonances on the water transport to objects in the circumprimary habitable zone of binary star systems}

\correspondingauthor{David Bancelin}
\email{david.bancelin@univie.ac.at}

\author[0000-0002-0786-7307]{David Bancelin}
\affil{Instit\"ut f\"ur Astrophysik \\
Universit\"at Wien, T\"urkenschanzstr. 17 \\
A-1180 Vienna, Austria}

\author{Elke Pilat-Lohinger}
\affiliation{Instit\"ut f\"ur Astrophysik \\
Universit\"at Wien, T\"urkenschanzstr. 17 \\
A-1180 Vienna, Austria}

\author{Thomas I. Maindl}
\affiliation{Instit\"ut f\"ur Astrophysik \\
Universit\"at Wien, T\"urkenschanzstr. 17 \\
A-1180 Vienna, Austria}

\author{Florian Ragossnig}
\affiliation{Instit\"ut f\"ur Astrophysik \\
Universit\"at Wien, T\"urkenschanzstr. 17 \\
A-1180 Vienna, Austria}

\author{Christoph Sch\"afer}
\affiliation{Instit\"ut f\"ur Astronomie und Astrophysik\\
Eberhard Karls Universit\"at T\"ubingen, Auf der	Morgenstelle 10 \\
D-72076 T\"ubingen, Germany}



\begin{abstract}

	We investigate the role of secular and mean motion resonances on the water transport from a belt of icy asteroids onto planets or embryos orbiting inside the circumprimary habitable zone (HZ) of a binary star system. In addition, the host-star has an accompanying gas giant planet. For a comparison, we perform two case studies where a secular resonance (SR) is located either inside the HZ close to 1.0 au (causing eccentric motion of a planet or embryos therein) or in the asteroid 
	belt, beyond the snow line. In the latter case, a higher flux of icy objects moving towards the HZ is expected. Collisions 
	between asteroids and objects in the HZ are treated analytically. Our purely dynamical study shows that the SR in the HZ boosts the water transport; however, collisions can occur at very high impact speeds. In this paper, we treat for the first time, realistic collisions using a GPU 3D-SPH code to assess the water loss in the projectile. Including the water loss into the dynamical results, we get more realistic values for the water mass fraction of the asteroid during an impact. We highlight that collisions occurring at high velocities greatly reduce the water content of the projectile and thus the amount of water transported to planets or embryos orbiting inside the HZ. Moreover, we discuss other effects that could modify our results, namely the asteroid's surface rate recession due to ice sublimation and the atmospheric drag contribution on the asteroids' mass loss. 

\end{abstract}




\section{Introduction}\label{S:intro}

For studies of Earth analogs in planetary systems, the existence of water on
such objects defines the habitability of this planet. The origin of water on
Earth is still an open question and subject of debates leading to
controversies. Endogenous and exogenous (asteroids, comets, and planetary embryos) were suggested separately to explain the water content of the 
Earth. It was believed that the water 
presently on Earth was carried by comets from the Oort cloud in the late veneer scenario \citep{owen95} or from the 
Jupiter family. However, both astronomical and geochemical analyses 
\citep{lecuyer98} and the recent data from Rosetta 
\citep{altwegg15} showed heterogeneous distributions of the D/H ratio compared to the 
Earth's. Since, models advocating 
impacts of large number of asteroids are favored.

It is well known that orbital resonances -- i.e. mean motion (MMR) and secular resonances (SR) -- play a key role in the architecture of a planetary system as shown in various studies for our solar system \citep{tsiganis05,gomes05,walsh11} and in binary systems \citep{mudryk06,thebault14}. Icy bodies trapped in orbital resonances are also 
potential water sources for objects orbiting inside the habitable zone (HZ). These water-rich bodies are either planetary embryos 
\citep{haghighipour07} or asteroids as shown for our Earth in \cite{morbidelli00} and \cite{obrien14}. More recently, 
\cite{bancelin15} gave a statistical estimate of the contribution of asteroids in bearing water material onto the circumprimary HZ of binary star systems. 
They showed that the differences in the statistical results strongly depend on the initial configuration of the binary 
star systems hosting a gas giant. \cite{bancelin16} showed that in such configurations an SR can be located within the planetesimal belt where an overlap with MMRs
may lead to strong perturbations causing a flux of asteroids moving towards
the HZ. In the case of icy asteroids, they will contribute to the water transport
onto a planet moving in the HZ. Studies by \cite{pilat15} and \cite{bazso15} introduced a
quick semi-analytical method to determine the location of the linear SR. This
method showed clearly how the position of the SR depends on the semi-major
axis, the eccentricity and the mass of both the giant planet and the secondary
star. Thus, in certain binary-planet configurations, the SR may help to
transport water into the HZ. While in other cases the highly
eccentric motion due to the SR  may influence the habitability of a planetary
system as it was shown for the solar system by \cite{pilat08b}, \cite{pilat08a} and \cite{pilat15b}.\\

The aim of the present study is, on one hand, to highlight in detail the influence of orbital resonances on the water transport by icy asteroids from a planetesimal belt to bigger objects (embryos or planets) orbiting the host-star within the HZ. On the other hand, most studies of planetary formation performed so far do not treat collisions properly. Indeed, collisions occur far from the real collision distance (generally at several Hill radii) and using a perfect merging approach. This in turn can result in overestimating the size and the water content of the body resulting from this collision. In this paper, we also aim to provide, for the first time, a comparison between a purely merging approach and more realistic collisions using a GPU 3D-SPH code to model the water transferred from a small projectile (asteroid) to a bigger target (Moon-, Mars-, and Earth-sized objects). \\

For various binary star-giant planet configurations, we investigate the two following scenarios.
\begin{itemize}
	\item [(1)] Eccentric planetary motion inside the HZ (induced by an SR at $\sim$ 1.0 au and/or MMRs) and an asteroid belt perturbed by MMRs.
	\item [(2)] Nearly circular planetary orbits inside the HZ and an asteroid belt 
	perturbed by an SR and MMRs.
\end{itemize}
In Sect. \ref{S:model}, we define our dynamical model for the various binary star systems, the gas giant and an icy asteroid belt. We also describe therein our algorithm for the collision assessment. The results are presented and analyzed in Sect. \ref{S:results}. Finally, we discuss in Sect. \ref{S:discussion} the possibility of processes causing water loss from an asteroid's surface,  e.g. ice sublimation and 
atmospheric pressure.

\section{Methods}\label{S:model}

\subsection{Binary stars and gas giant configurations}

We focused our work on binary star systems with two G-type stars with mass equal to 1$M_{\scriptscriptstyle \odot}$. 
The initial orbital separations are $a_{\scriptscriptstyle \text{b}}$ = 50 au or 100 au and the binary's eccentricity is 
$e_{\scriptscriptstyle \text{b}}$ = 0.1 or 0.3. Since we study the planar case, all initial inclinations are set to 0$^{\circ}$. Our studied systems host a gas 
giant planet initially moving on a circular orbit with a mass equal to the 
mass of Jupiter. As pointed out by \cite{pilat15}, the location of the SR depends both on the orbital 
elements of the secondary and the gas giant. Since we fixed the binary's semi-major axis to 50 and 100 au for our numerical study, the position of the gas giant ($a_{\scriptscriptstyle \text{GG}}$)
changes for the different systems depending on whether we analyze the perturbation of the SR in the HZ or how the SR affects the water transport. The latter can be considered as a positive influence on the 
habitability because the SR is located in the planetesimal belt, which is beyond 2.7 au. With the conditions that the SR should either be beyond 2.7 au or around 1.0 au, we determine the according semi-major axis of the giant planet using the semi-analytical method of \cite{pilat15} and \cite{bazso15}. We summarize the values of $a_{\scriptscriptstyle \text{GG}}$ in Tab. \ref{T:agg}.
\begin{table}[!h]
	\begin{center}
		\caption{Gas giant's semi-major axis $a_{\scriptscriptstyle \text{GG}}$ according to the binary's periapsis 
			distance $q_{\scriptscriptstyle \text{b}}$. Second and third columns are values 
			of $a_{\scriptscriptstyle \text{GG}}$ 
			such as the secular resonance (SR) lies inside the HZ or inside the asteroid belt, 
			respectively.}
		\label{T:agg}
		\begin{tabular}{ccc}
			\hline
			$q_{\scriptscriptstyle \text{b}}$ (au) & $a_{\scriptscriptstyle \text{GG}}$ (au) & $a_{\scriptscriptstyle \text{GG}}$ 
			(au) \\
			& (SR inside HZ) & (SR inside the disk)\\
			\hline
			35 & 3.00 & 5.2 \cr
			\hline
			45 & 3.10 & 5.2 \cr
			\hline 
			70 & 4.75 & 7.2\cr
			\hline
			90 & 4.5 & 7.2 \cr
			
			\hline
		\end{tabular}
	\end{center}
\end{table}

\subsection{Icy Planetesimals Disk Modeling}

We modeled a disk of 1000 icy Ceres-like planetesimals with mass $M_{\scriptscriptstyle \text{CERES}} = 4.76\,\times 10^{\scriptscriptstyle -10}\,M_{\odot}$ inside and beyond the 
orbit of the gas giant. They have initially nearly circular and planar orbits with initial eccentricities and inclinations randomly chosen below 0.01 and 1$^\circ$, respectively. To avoid strong 
initial interactions with the gas giant, we assumed that it has gravitationally cleared a path in the disk around its 
orbit. We defined the width of this path as $\pm 3\,R_{\scriptscriptstyle {\text{H},\text{GG}}}$, where 
$R_{\scriptscriptstyle {\text{H},\text{GG}}}$ is the giant planet's Hill radius. The small bodies' semi-major axis $a_{\scriptscriptstyle {\text{A}}}$ is uniformly distributed into three regions from an inner  to an outer border, the latter being defined 
by the stability criteria \citep{holman99, pilat02}. The three regions are defined as follows.
\begin{itemize}
	\item [1.] $\mathcal{R}_{1}$ is for 2.0 $\le$ $a_{\scriptscriptstyle \text{A}}$ $\le$ 2.7 au,
	\item [2.] $\mathcal{R}_{2}$ is for 2.7 $\le$ $a_{\scriptscriptstyle \text{A}}$ $\le$ $a_{\scriptscriptstyle {\text {GG}}} - 3\,R_{\scriptscriptstyle {\text{H},\text{GG}}}$,
	\item [3.] $\mathcal{R}_{3}$ is for $a_{\scriptscriptstyle \text{A}}$ $\ge$ $a_{\scriptscriptstyle {\text {GG}}} +
	3\,R_{\scriptscriptstyle {\text{H},\text{GG}}}$
\end{itemize}
Because of this definition, region $\mathcal{R}_{2}$ does not exist for the binary with $q_{\scriptscriptstyle \text{b}}$ = 35 au and 45 au as $a_{\scriptscriptstyle {\text {GG}}} - 3\,R_{\scriptscriptstyle {\text{H},\text{GG}}} < 2.7$ au. 

Each asteroid is assigned a water mass fraction (hereafter wmf) using a linear approximation and with borders defined in the following way. 
\begin{itemize}
	\item [1.] $a_{\scriptscriptstyle \text{A}} \in\,\mathcal{R}_{1}$, 1 $\le$ wmf $\le$ 10\%
	\item [2.] $a_{\scriptscriptstyle \text{A}} \in\,\mathcal{R}_{2}$, 10 $\le$ wmf $\le$ 15\%
	\item [3.] $a_{\scriptscriptstyle \text{A}} \in\,\mathcal{R}_{3}$, wmf = 20\%
\end{itemize}

Because the distribution of $a_{\scriptscriptstyle \text{A}}$ within the disk is different for the various binaries investigated (as it depends on the location of the gas giant) the total amount of water $\mathcal{W}_{\scriptscriptstyle \text{TOT}}$ (expressed in Earth-ocean unit\footnote{one Earth-ocean weights $1.5\,\times 10^{21}$ kg of H$_2$O}) contained in the disk also varies, as shown in Tab. \ref{T:tot_water}.

\begin{table}[!h]
	\begin{center}
		\caption{Total amount of water $\mathcal{W}_{\scriptscriptstyle \text{TOT}}$ (expressed in Earth-ocean unit) contained in the icy asteroids belt.}
		\label{T:tot_water}
		\begin{tabular}{ccc}
			\hline
			$q_{\scriptscriptstyle \text{b}}$ (au) & $\mathcal{W}_{\scriptscriptstyle \text{TOT}}$ (Earth-ocean) &  $\mathcal{W}_{\scriptscriptstyle \text{TOT}}$ (Earth-ocean) \\
			& (SR inside HZ) & (SR inside the disk)\\
			\hline
			35 & 58.6 & 78.2 \cr
			\hline
			45 & 60.2 & 78.2 \cr
			\hline 
			70 & 75.6 & 81.5\cr
			\hline
			90 & 76.7 & 81.5 \cr
			
			\hline
		\end{tabular}
	\end{center}
\end{table}

\subsection{Impact modeling}\label{S:water}

\begin{figure}
	\centering{\includegraphics[angle=-90,width=\columnwidth]{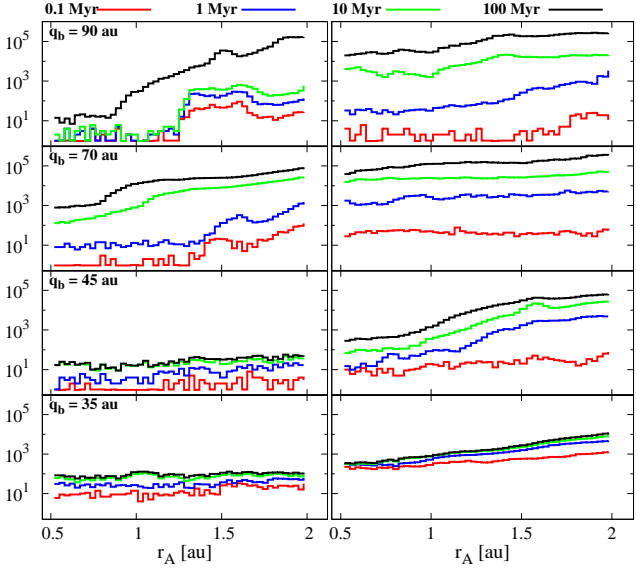}}
	\caption{Cumulative distribution of $r_{\scriptscriptstyle \text{A}}$ for asteroids reaching distances $<$ 2.0 au. The color lines indicate different integration times. The panels show the results for the studied $q_{\scriptscriptstyle \text{b}}$ (increasing $q_{\scriptscriptstyle \text{b}}$ from bottom to top. Left and right panels are for the SR lying inside or outside the HZ, respectively.}\label{F:HZc}
\end{figure}

The minimum orbital intersection distance (MOID) is widely used in celestial mechanics to determine whether an object can be a potential danger for the Earth. The MOID corresponds to the closest distance between two 
keplerian orbits, regardless their real position on their respective 
trajectories. For our study, we need to derive the MOID in order to simulate impacts between icy objects initially located in the asteroid belt and 
embryos-to-planet-sized bodies (EPs) initially inside the HZ.  They are either 
Moon- and Mars-sized (embryos) or Earth-sized (planets) (see Table \ref{T:planet} for the physical mass and radius). For this purpose, we performed for both scenarios (i.e. the SR either inside the HZ or inside the asteroid belt) two separate simulations: considering the gravitational perturbations of the stars and the gas giant, we integrated separately for 100 Myr using the Radau integrator.
\begin{itemize}
	\item [(1)] The asteroid belt to assess the orbital distribution of asteroids reaching distances $r_{\scriptscriptstyle \text{A}} < 2.0$ au 
	to the primary star. Figure \ref{F:HZc} shows the cumulative distribution of the number of crossings as a function of $r_{\scriptscriptstyle \text{A}}$ au for the different binary star systems investigated and for several integration times. The left panels show results for a SR lying inside the HZ and the right panels are the results for the SR in the asteroid belt. As indicated by the different colors, one can see that in both cases, asteroids can rapidly 
	increase their eccentricity and reach Mercury's distances within 1 Myr (blue 
	histograms). We can also notice that the number of crossings is higher when the SR is located 
	inside the belt (right panels). This was also pointed out by \cite{bancelin16} and can be explained by possible overlaps of the SR with MMRs.\\
	\item [(2)] EPs at initial location $a_{\scriptscriptstyle {\text{EP}}}$ to assess the evolution of their eccentricity with time. They are uniformly distributed over 48 positions within the HZ (with borders defined according to \cite{kopparapu13}) and initially move in circular orbits.
\end{itemize}
\begin{table}[!h]
	\begin{center}
		\caption{Physical parameters, mass $M_{\scriptscriptstyle \text{EP}}$ and radius $R_{\scriptscriptstyle \text{EP}}$, 
			of the EPs considered in our simulations expressed in 
			Earth-mass (M$_{\oplus}$ = 3.0$\times$10$^{-6}$ M$_{\odot}$) and Earth-radius (R$_{\oplus}$ = 6378.0 km)}
		\label{T:planet}
		\begin{tabular}{ccc}
			\hline
			& Mass $M_{\scriptscriptstyle \text{EP}}$ & Radius $R_{\scriptscriptstyle \text{EP}}$ \\
			& (M$_{\scriptscriptstyle \oplus}$) & (R$_{\scriptscriptstyle \oplus}$) \\
			\hline
			Earth & 1.000 & 1.00 \cr
			\hline
			Mars & 0.100 & 0.53 \cr
			\hline 
			Moon & 0.012 & 0.27 \cr
			
			\hline
		\end{tabular}
	\end{center}
\end{table}

We combined the results of these two integrations to analytically compute the MOID using the algorithm described in \cite{sitarski68}. 
We define a collision if the MOID is comparable to the EP's radius $R_{\scriptscriptstyle \text{EP}}$. For the collision assessment, we use the following  five-steps algorithm.
\begin{itemize}
	\item [(1)] For a given position $a_{\scriptscriptstyle \text{EP}}$, we check if $q_{\scriptscriptstyle \text{EP}}$ 
	$\le$ 
	$r_{\scriptscriptstyle \text{A}}$ $\le$ $Q_{\scriptscriptstyle \text{EP}}$ where $q_{\scriptscriptstyle \text{EP}}$ and 
	$Q_{\scriptscriptstyle \text{EP}}$ are the periapsis and apoapsis 
	distances of the EPs respectively which are defined according to 
	$e_{\scriptscriptstyle \text{EP}}$ = $e_{\scriptscriptstyle \text{max}} (t)$ with 
	$e_{\scriptscriptstyle \text{max}} (t)$ the maximum of the EP's eccentricity for 
	different 
	period of integration. If this condition is fulfilled, then we go to the next step.\\
	
	\item [(2)] In reality, $e_{\scriptscriptstyle \text{EP}}$ has periodic variations between 0 (its initial value) and 
	$e_{\scriptscriptstyle 
		\text{max}} (t)$. Thus, we define the function $Y(e_{\scriptscriptstyle \text{EP}})$ = 
	$d_{\scriptscriptstyle \text{min}} 
	(e_{\scriptscriptstyle \text{EP}})$ - $R_{\scriptscriptstyle \text{EP}}$ where $d_{\scriptscriptstyle \text{min}}$ is 
	the MOID. According to the sign of the product $Y(e_{\scriptscriptstyle \text{EP}} = 0)$ $\times$ 
	$Y(e_{\scriptscriptstyle \text{EP}} = e_{\scriptscriptstyle \text{max}})$ we use different procedures: if the sign 
	is negative then we use a 
	regula falsi procedure in order to find the value of $e_{\scriptscriptstyle \text{EP}}$ giving $Y(e_{\scriptscriptstyle 
		\text{EP}}) \approx 0$. If the sign is positive, then we use a dichotomy 
	procedure to find a value 
	of $e_{\scriptscriptstyle \text{EP}}$ leading to an impact.\\
	
	\item [(3)] When a collision is found, i.e. if $d_{\scriptscriptstyle {\text{min}}}\,\le\,R_{\scriptscriptstyle 
		{\text{EP}}}$ we also derive, from the MOID, the true anomalies of the EP and 
	the asteroid in order to 
	compute the relative impact velocity and impact angle of the asteroid.\\
	
	\item [(4)] If the previous condition is fulfilled, then we define:
	\begin{equation*}
	\left\{ 
	\begin{array}{l}
	\mathcal{W}_{\scriptscriptstyle \text{k}}(a_{\scriptscriptstyle \text{EP}}, t) = \displaystyle \frac{M_{\scriptscriptstyle \text{CERES}}\, \times wmf_{\scriptscriptstyle \text{k}}}{M_{\scriptscriptstyle \text{H}_{\scriptscriptstyle 2} \text{O}}} \\
	
	\displaystyle \widetilde{\mathcal{W}}_{\scriptscriptstyle \text{k}}(a_{\scriptscriptstyle \text{EP}}, t) = \frac{1}{N_{\scriptscriptstyle \text{k}}}  \displaystyle \sum_{k=1}^{N_{\scriptscriptstyle \text{k}}} \frac{M_{\scriptscriptstyle \text{CERES}}\, \times wmf_{\scriptscriptstyle \text{k}}\times\,\left( 1-\omega_{\scriptscriptstyle \text{c}} (a_{\scriptscriptstyle \text{EP}}) \right)}{M_{\scriptscriptstyle \text{H}_{\scriptscriptstyle 2} \text{O}}}
	\end{array}
	\right.
	\end{equation*}
	
 as the quantity of water (in Earth-ocean unit) delivered by an asteroid number $k$ (with water mass fraction $wmf_{\scriptscriptstyle \text{k}}$), at an intermediate integration time $t$, to the EP initially at a semi-major axis $a_{\scriptscriptstyle \text{EP}}$. Here, $M_{\scriptscriptstyle \text{H}_{\scriptscriptstyle 2} \text{O}} = 1.5\,\times 10^{21}$ kg of H$_2$O is the mass of one Earth-ocean. The first term $\mathcal{W}_{\scriptscriptstyle \text{k}}$ assumes a merging approach in which the whole water content of the asteroid is delivered to the EP without assuming any water loss processes. The second term $\displaystyle \widetilde{\mathcal{W}}_{\scriptscriptstyle \text{k}}$ takes into account a water loss factor $\omega_{\scriptscriptstyle \text{c}}$ induced by a water loss mechanism. Here, $N_{\scriptscriptstyle \text{k}}$ is the number of possible collisions of the asteroid $k$.\\
	
	\item [(5)] At the end of the procedure, we can derive the total fraction of water delivered to the EP: $\mathcal{W}_{\scriptscriptstyle \text{EP}} (a_{\scriptscriptstyle \text{EP}},t) = \displaystyle \frac{1}{\mathcal{W}_{\scriptscriptstyle \text{TOT}}} \sum_{k=1}^{n_{\scriptscriptstyle \text{i}}}\mathcal{W}_{\scriptscriptstyle \text{k}}(a_{\scriptscriptstyle \text{EP}},t)$ and $\mathcal{\widetilde{W}}_{\scriptscriptstyle \text{EP}} (a_{\scriptscriptstyle \text{EP}},t) = \displaystyle \frac{1}{\mathcal{W}_{\scriptscriptstyle \text{TOT}}} \sum_{k=1}^{n_{\scriptscriptstyle \text{i}}}\mathcal{\widetilde{W}}_{\scriptscriptstyle \text{k}}(a_{\scriptscriptstyle \text{EP}},t)$ where $n_{\scriptscriptstyle \text{i}}$ is the number of impactors. We also compute the median values of the asteroids' impact velocities $\overline{v}_{\scriptscriptstyle \text{i}}(a_{\scriptscriptstyle \text{EP}})$ and angles $\overline{\theta}_{\scriptscriptstyle \text{i}}(a_{\scriptscriptstyle \text{EP}})$ according to the total number of possible collisions found.

\end{itemize}

\section{Results}\label{S:results}
\begin{figure*} 
	\centering{
		\begin{tabular}{cc}
			\includegraphics[angle=-90,width=0.485\textwidth]{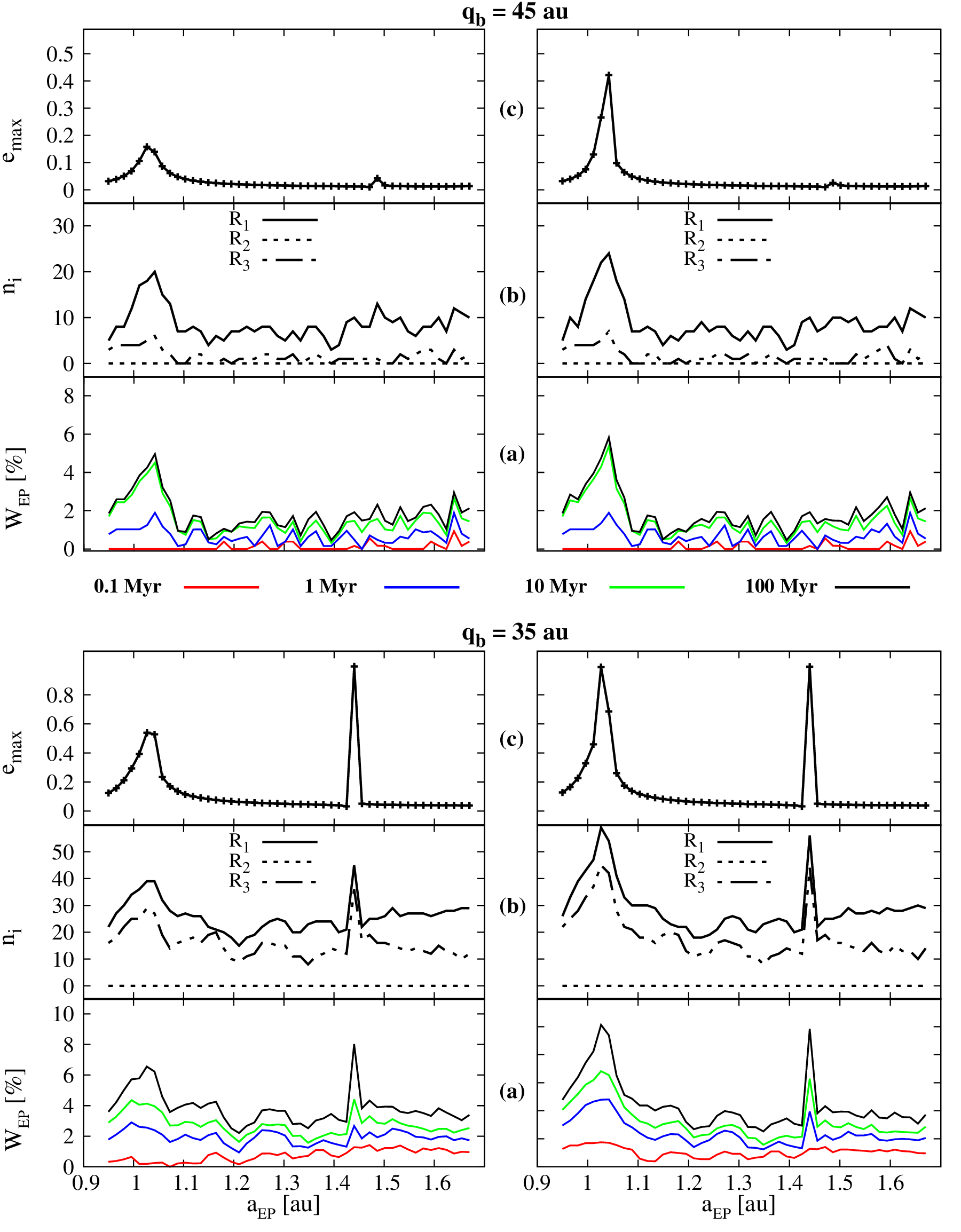} &
			\includegraphics[angle=-90,width=0.485\textwidth]{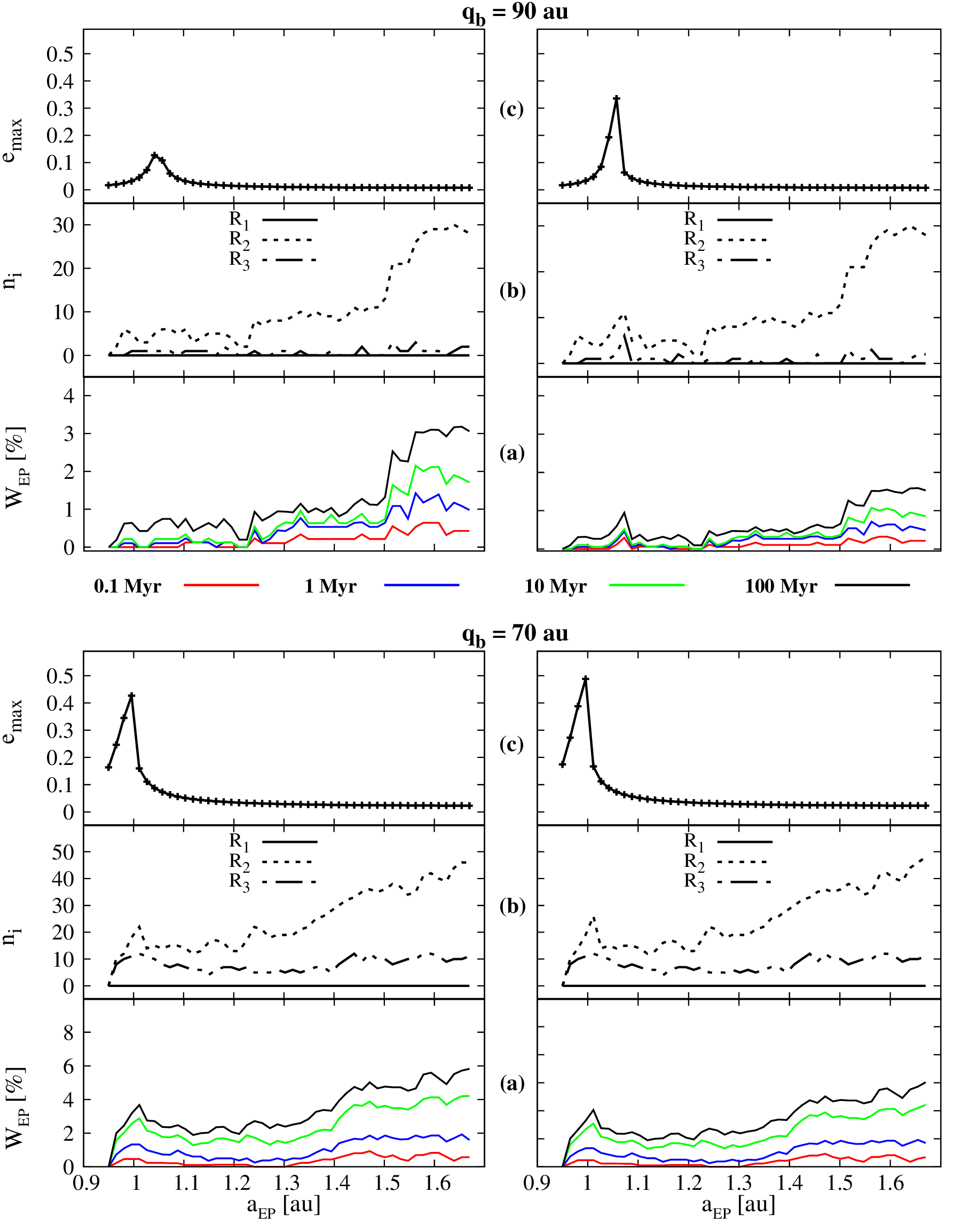}\\
	\end{tabular}}
	\caption{\textit{Left} and \textit{right} panels: results for $a_{\scriptscriptstyle \text{b}}$ = 50 au 
		and 100 au respectively. For a given value of $q_{\scriptscriptstyle \text{b}}$, the \textit{left} and
		\textit{right sub-panels} show results for an Earth- and Mars/Moon-like objects respectively. For each object, from bottom to 
		top, are shown in (a) the fraction of incoming water  $\mathcal{W}_{\scriptscriptstyle \text{EP}}$ at the EP's surface for several timescales (0,1 Myr in red, 1 Myr in blue, 10 Myr in green and 100 Myr in black); in (b) the total number of impactors $n_{\scriptscriptstyle \text{i}}$ (up to 100 Myr) for each initial region $\mathcal{R}_{1}$ (solid line), $\mathcal{R}_{2}$ (dashed line) and $\mathcal{R}_{3}$ (dotted-solid line); in (c) the maximum eccentricity $e_{\scriptscriptstyle \text{max}}$ (up to 100 Myr).}\label{F:water_all}
\end{figure*}

\begin{figure*} 
	\centering{
		\includegraphics[angle=-90,width=\textwidth]{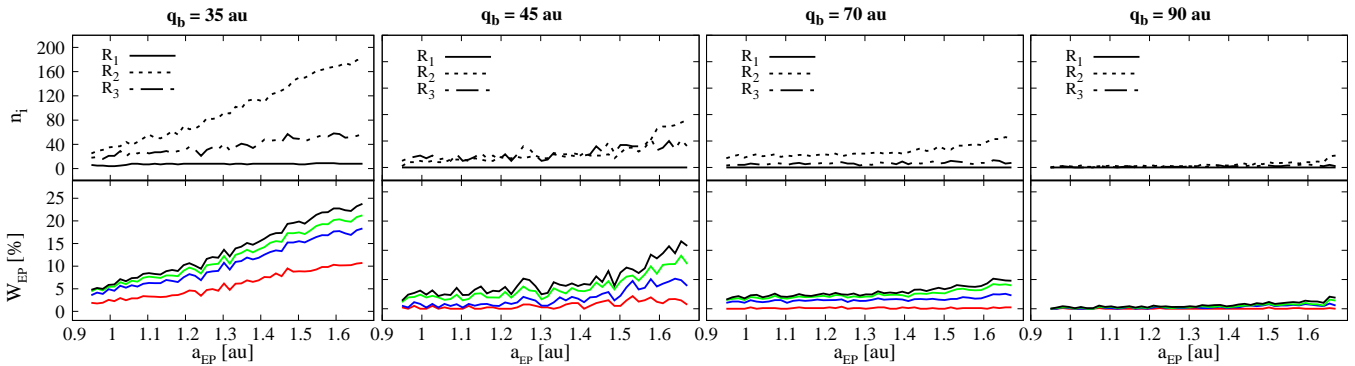}}
	\caption{Same as in Fig. \ref{F:water_all} but for the case of an SR inside the asteroid belt. From \textit{left} to 
		\textit{right panels} are shown results for increasing value of $q_{\scriptscriptstyle \text{b}}$. As the eccentricity evolution is the same for each EP, only one of the EP's results is displayed for $\mathcal{W}_{\scriptscriptstyle \text{EP}}$ (\textit{bottom panel}) and 
		$n_{\scriptscriptstyle \text{i}}$ (\textit{top panel}).}\label{F:belt}
\end{figure*}

In Fig. \ref{F:water_all}, we show detailed results obtained when a SR is inside the 
HZ for a secondary either at $a_{\scriptscriptstyle \text{b}}$ = 50 au (left panels) or at $a_{\scriptscriptstyle \text{b}}$ = 100 au (right panels). 
For each binary's periapsis distance we show the outcome for an Earth- (left sub-panels) and Mars/Moon-sized bodies (right sub-panels) as a function of their initial semi-major axis. From bottom to top, the panels represent in (a) the cumulative fraction of incoming water $\mathcal{W}_{\scriptscriptstyle \text{EP}} (a_{\scriptscriptstyle \text{EP}},t)$ on the EP (for intermediate integration times $t$ = 0.1 -- 1 -- 10 -- 100 Myr in red, blue, green an solid black lines respectively); (b) the number of asteroids $n_{\scriptscriptstyle \text{i}}$ impacting the EP's surface (at $t$ = 100 Myr); (c) the maximum eccentricity $e_{\scriptscriptstyle \text{max}}$ reached by an EP (over $t$ = 100 Myr). Similar results are presented in Fig. \ref{F:belt} for the case of a SR inside the asteroid belt. We did not represent the EP's  eccentricity distribution in the HZ like in Fig. \ref{F:HZc} because their orbit remains almost circular during the integration time, whatever the size of the EP. Besides, only results for one of the EPs are displayed.

\subsection{Water transport: merging approach}\label{SS:merging}

In Fig. \ref{F:water_all}, one can recognize in sub-panels (c) the SR around 1.0 au causing a relatively high eccentric motion with decreasing $q_{\scriptscriptstyle \text{b}}$ and $M_{\scriptscriptstyle \text{EP}}$. Due to the proximity of the gas giant, we can also notice, for $q_{\scriptscriptstyle \text{b}}$ = 35 au, the presence of the 3:1 MMR at 1.44 au also causing high eccentric motions in this area. 
As a consequence, we can see for $a_{\scriptscriptstyle \text{b}}$ = 50 au (left panels and sub-panels (b)) that an EP initially moving inside an orbital resonance can collide with more asteroids than if they were orbiting outside the SR or MMR. Both regions $\mathcal{R}_{1}$ and $\mathcal{R}_{3}$ can be water sources for EPs within the entire HZ. Indeed, as shown in the sub-panels (a), within the first 1 Myr, the quantity of water $\mathcal{W}_{\scriptscriptstyle \text{EP}}$ delivered to an EP, initially located inside an orbital resonance, is boosted as it can encounter more asteroids, not only those crossing the HZ but also those with distance 
$r_{\scriptscriptstyle \text{A}}$ beyond the HZ (see Fig. \ref{F:HZc}). For $a_{\scriptscriptstyle \text{b}}$ = 100 au (right panels), despite the relatively high eccentricity of EPs around 1.0 au, $n_{\scriptscriptstyle \text{i}}$ is mainly maximum for $a_{\scriptscriptstyle \text{EP}}$ close to the outer border of the HZ, following the $r_{\scriptscriptstyle \text{A}}$ distribution (see Fig. \ref{F:HZc}). Therefore, as shown in sub-panels (a), EPs at 1.0 can attract equal or even less water than EPs outside the SR. 
For both values of $a_{\scriptscriptstyle \text{b}}$ investigated, we can notice that $n_{\scriptscriptstyle \text{i}}$ represents only a small fraction of the total number of asteroids in the disk (less than 10\%). Also, $\mathcal{W}_{\scriptscriptstyle \text{EP}}$ can reach a maximum of 9\% of the total amount of water in the disk.

Figure \ref{F:belt} illustrates results for the second case (SR inside the belt). 
Because of the huge number of impactors, up to four times higher ($\sim$ 25\% of the initial population for instance for $q_{\scriptscriptstyle \text{b}}$ = 35 au) than if a SR was inside the HZ, the water transport to the HZ is boosted and $\mathcal{W}_{\scriptscriptstyle \text{EP}}\,\sim\,15\%$ for $q_{\scriptscriptstyle \text{b}}$ = 45 au and even $\sim\,25\%$ for $q_{\scriptscriptstyle \text{b}}$ = 35 au. For each binary configuration studied, the main source of impactors is located in the region $\mathcal{R}_{\scriptscriptstyle 2}$. As already proved by \cite{bancelin16}, this region is the primary source of icy asteroids reaching the HZ for this configuration, as both orbital resonances overlap in the asteroid belt. Contrary to the previous case (SR $\in$ HZ), the water distribution in the HZ follows the $r_{\scriptscriptstyle \text{A}}$ distribution and EPs located around 1.0 au, on circular motion, can interact with fewer icy asteroids than EPs located at larger semi major axes. Therefore, only a small fraction of the $\mathcal{W}_{\scriptscriptstyle \text{TOT}}$, less than 5\%, can be transported to an EP.

\begin{figure*}
	\centering{
		\includegraphics[angle=-90,width=\textwidth]{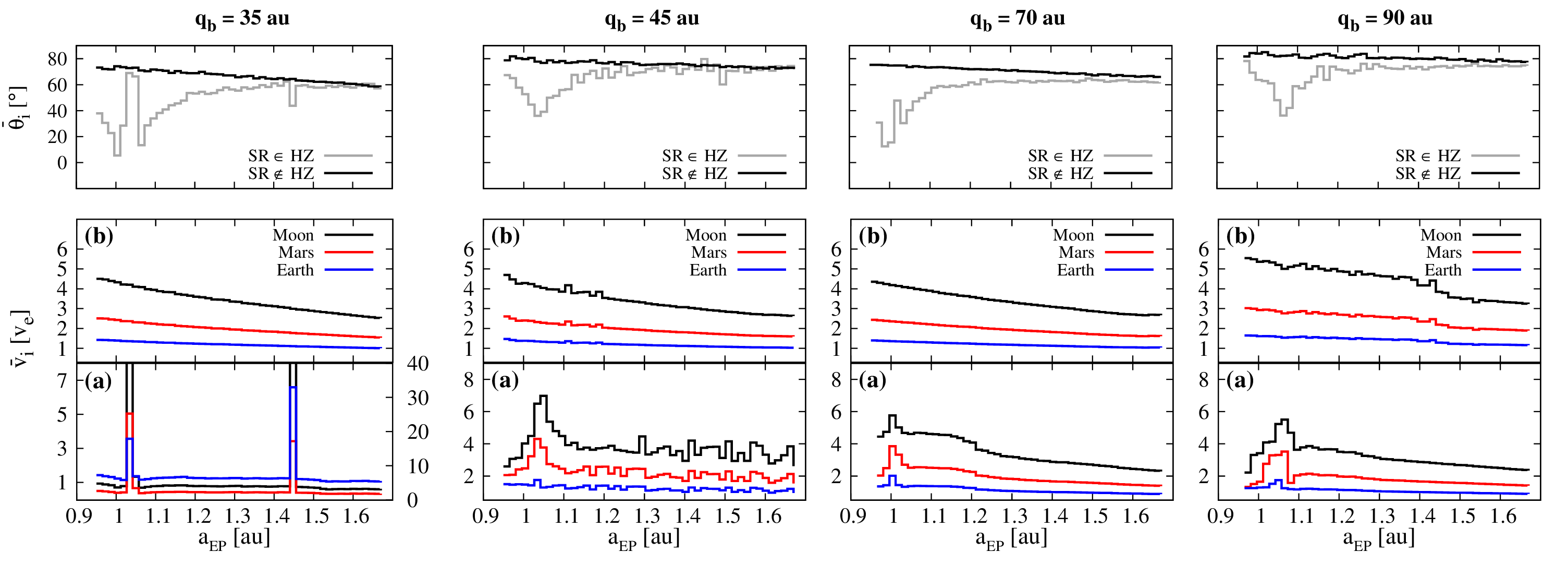}}
	\caption{Impact angles $\overline{\theta}_{\scriptscriptstyle \text{i}}$ velocities $\overline{v}_{\scriptscriptstyle \text{i}}$ (\textit{top} and \textit{bottom} panels respectively) with respect to $a_{\scriptscriptstyle \text{EP}}$ and for increasing values of $q_{\scriptscriptstyle \text{b}}$ from left to right. Colors in the \textit{bottom} panels are for impact velocities at the surface of Earth, Mars and Moon objects (solid blue, red and black lines respectively) depending on whether SR lies inside or outside the HZ (panels \textit{a} and \textit{b} respectively). For $q_{\scriptscriptstyle \text{b}}$ = 35 au, values of $\overline{v}_{\scriptscriptstyle \text{i}}$ for Moon/Mars have to be read on the right y-axis as the values are much larger than the Earth (left y-axis). Colors in the \textit{top} panel are for SR inside the HZ (solid grey line) or inside the belt (solid black line).}\label{F:impact}
	
\end{figure*}

\subsection{Impact parameters statistics}\label{SS:vel}

In Fig. \ref{F:impact}, we represent the impact angles and velocities distribution $\overline{\theta}_{\scriptscriptstyle \text{i}}$ and $\overline{v}_{\scriptscriptstyle \text{i}}$ (top and bottom panels respectively). Results are compared for both locations of the SR: inside the HZ (grey solid line for $\overline{\theta}_{\scriptscriptstyle \text{i}}$ and panels (a) for $\overline{v}_{\scriptscriptstyle \text{i}}$) and outside the HZ (black solid line for $\overline{\theta}_{\scriptscriptstyle \text{i}}$ and panels (b) for $\overline{v}_{\scriptscriptstyle \text{i}}$). In the bottom panels, we distinguished impact velocities on Earth-, Mars- and Moon-sized objects (solid blue, red and black lines respectively).
For both cases (SR inside and outside the HZ), the impact velocity distribution 
inside the HZ shows different modes depending on $a_{\scriptscriptstyle \text{EP}}$: outside the orbital resonances, statistically, we have for asteroids impacting an 
Earth-sized $1.0 < \overline{v}_{\scriptscriptstyle \text{i}} < 1.5\,v_{\scriptscriptstyle \text{e}}$, a Mars-sized $1.5< \overline{v}_{\scriptscriptstyle \text{i}} < 
2.5\,v_{\scriptscriptstyle \text{e}}$ and a Moon-sized $2.5 < \overline{v}_{\scriptscriptstyle \text{i}} < 6.0\,v_{\scriptscriptstyle \text{e}}$ (all 
expressed in escape velocity 
$v_{\scriptscriptstyle \text{e}}$ units of the respective EP). However, near an orbital resonance, 
$\overline{v}_{\scriptscriptstyle \text{i}}$ can be slightly higher as seen in Fig. 
\ref{F:impact}. For instance, for $q_{\scriptscriptstyle \text{b}}$ = 35 au (for this particular system, one should read results for the Earth case on the left $y$-axis and for the Mars/Moon case on the right $y$-axis), $\overline{v}_{\scriptscriptstyle \text{i}}$ can reach up to 6.0$\,v_{\scriptscriptstyle \text{e}}$ for objects impacting an Earth-, 25$\,v_{\scriptscriptstyle \text{e}}$ for a Mars- and beyond 
40$\,v_{\scriptscriptstyle \text{e}}$ for a Moon-sized body. In a similar way, $\overline{\theta}_{\scriptscriptstyle \text{i}}$ has high variations between 20$^\circ$ 
and 80$^\circ$ mainly near the SR. However, when an EP is not initially located inside an orbital resonance (both Fig. 
\ref{F:water_all} and \ref{F:belt}), $\overline{\theta}_{\scriptscriptstyle \text{i}}$ is in average above 50$^\circ$.\\

\subsection{Real collision modeling and water loss}\label{SS:SPH}

A key parameter of water and mass loss is the collision modeling itself when the asteroid arrives at the EP's 
surface. To verify the validity of the perfect merging assumption regarding collisional water loss, we performed simulations of water-rich Ceres-sized asteroids with dry 
targets in the mass range $M_{\scriptscriptstyle \mathrm{EP}}\/$ equal to 1\,$M_{\scriptscriptstyle \mathrm{MOON}}\/$, 
1\,$M_{\scriptscriptstyle \mathrm{MARS}}\/$ and 1\,$M_{\scriptscriptstyle \mathrm{EARTH}}\/$. We assume the Ceres-like 
impactor with $wmf = 15\%$ water content in a mantle around a rocky core. 
The simulations are performed with our parallel (GPU) 3D smooth particle hydrodynamics (SPH) code introduced by 
\cite{maisch13} and \cite{schrie16}. It includes self-gravity and implements solid-body continuum mechanics with a 
damage model for fracture and brittle failure along the lines of \cite{benasp94,benasp95}. The material behavior is 
modeled using the Tillotson equation of state \citep{til62}. A tensorial correction \citep{schspe07} warrants zeroth order consistency.

As we want to focus on merging, especially to verify a perfect merging assumption, we chose a collision angle leading to a merging outcome. Because larger angles result in hit-and-run configurations \citep{leinhardt12,maindl14}, we chose a smaller angle in order to ensure a merging configuration. We simulated impacts on targets with different sizes at an impact angle of $30^\circ\/$ 
with initial collision velocities (taken from Fig. \ref{F:impact}) $\overline{v}_{\scriptscriptstyle \text{i}}$ = 2; 5 and 30\,$v_{\scriptscriptstyle \text{e}}$ 
for the Moon,  $\overline{v}_{\scriptscriptstyle \text{i}}$ = 1; 5 and 20\,$v_{\scriptscriptstyle \text{e}}$ for Mars and 
$\overline{v}_{\scriptscriptstyle \text{i}}$ = 1; 3 and 5\,$v_{\scriptscriptstyle 
	\text{e}}$ for the Earth. Except for this latter 
case ($\overline{v}_{\scriptscriptstyle \text{i}}=5\,v_{\scriptscriptstyle \text{e} }$) 
for which about 1 million SPH particles are used,  most of the scenarios are 
resolved in about 500,000 SPH 
particles. All objects were relaxed by initializing them with self-consistent hydrostatic structures, and internal energies following adiabatic compression as described in \cite{burger15}.

In order to estimate an upper limit for water retained on the surviving body after the collision, we investigate the 
dynamics of matter ejected from the impact site\footnote{Note that we use the term \emph{ejecta\/} for material 
	accelerated away from the impact site regardless whether it originates from the projectile or the target.}. Collision 
fragments moving faster than the escape velocity are considered lost. This approach does not take into account losses 
due to (a) sublimation depending on the radiative environment of the individual collision and/or heating due to the 
impact process itself or (b) due to the interaction with an atmosphere that may survive the impact; we will model both 
aspects in a subsequent study. The water loss results therefore represent a lower loss limit.
\begin{figure}
	{\includegraphics[width=\columnwidth]{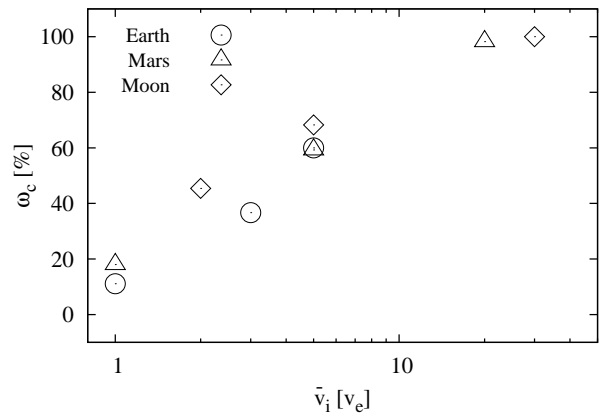}}
	\caption{Water loss $\omega_{\scriptscriptstyle \text{c}}$ as a mass fraction of the initial total water after impacts at velocity $\overline{v}_\mathrm{i}\/$ in 
		units of the EP's escape velocity $v_\mathrm{e}\/$.}
	\label{fig:wloss}
\end{figure}
Figure~\ref{fig:wloss} summarizes the water loss $\omega_{\scriptscriptstyle \text{c}}$ in the collision scenarios. All but the most extreme scenario result in a merged main survivor retaining most of the mass. The exception is the $30\,v_\mathrm{e}\/$ impact 
of a Ceres-like body onto the Moon which leads to mutual destruction of the bodies into a debris cloud and hence a loss 
of all volatile constituents. The other very fast impact (Mars at $20\,v_\mathrm{e}\/$) results in a merged survivor 
that retains just under 2\,wt-\% of the available water. For the lower collision velocities we observe water loss rates 
between 11\,wt-\% and 68\,wt-\%. This demonstrates that even with an approximate model, there are significantly different results compared to the perfect merging assumption. If plotted versus the impact velocity in terms of the mutual escape velocity as in 
Fig.~\ref{fig:wloss}, there is a strong correlation with the impact velocity but only weak dependence on the absolute 
mass. \\

\begin{figure*} 
\centering{
	{\includegraphics[width=\columnwidth, angle=-90]{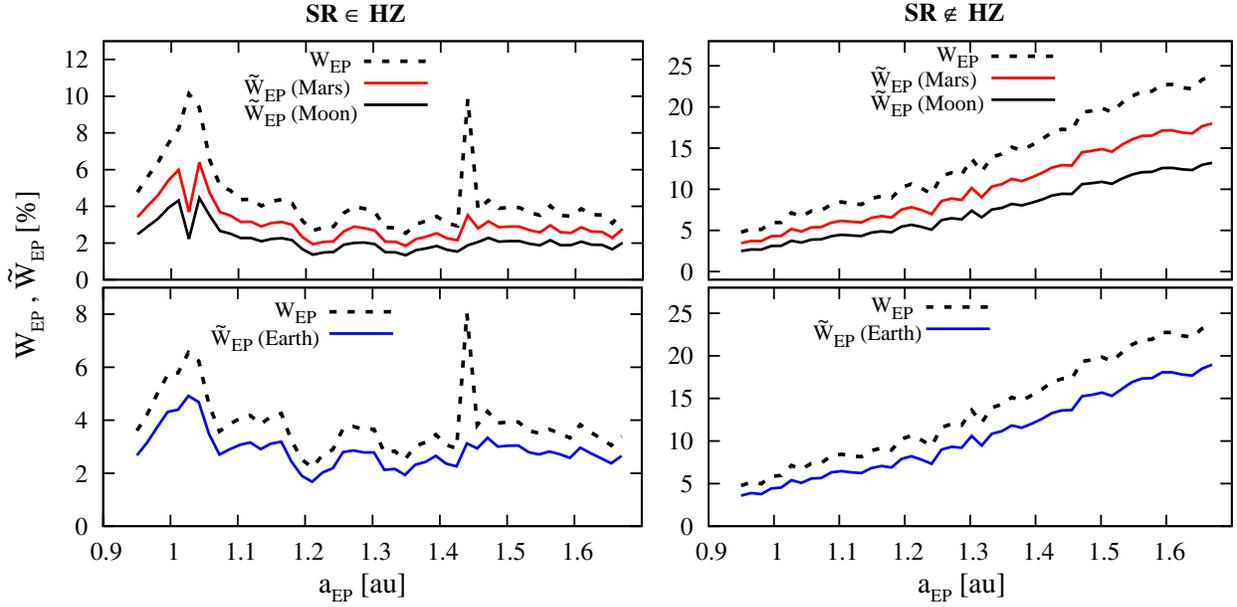}}}
	\caption{Fraction of incoming water with (solid lines) and without (dotted lines) our SPH collision model. \textit{Left} and \textit{right} panels are for SR inside the HZ and inside the asteroid belt, respectively.}
	\label{F:wloss_collision}
\end{figure*}

\subsection{Water transport due to real collisions}\label{SS:water_loss}

For a more realistic estimation of $\mathcal{W}_{\scriptscriptstyle \text{EP}}$, we use the previous results  shown in Fig. \ref{F:water_all} and \ref{F:belt} and take into account the fraction of water loss $\omega_{\scriptscriptstyle \text{c}}$ due to the physical impact between the asteroid and the EP of Fig. \ref{fig:wloss}. Using a linear extrapolation of $\omega_{\scriptscriptstyle \text {c}}$ between the minimum and maximum impact velocities derived for each EP, we are able to provide better estimates for the water transport. In Fig. \ref{F:wloss_collision}, we compare the fraction of water reaching the EP's surface with and without taking into account our study of SPH collisions represented by the solid and dotted lines respectively, i.e. $\mathcal{\widetilde{W}}_{\scriptscriptstyle \text{EP}}$ and $\mathcal{W}_{\scriptscriptstyle \text{EP}}$. The two top panels show results for Moon- and Mars-sized objects and the bottom panels, for Earth-sized bodies. We show the comparison for a binary with $q_{\scriptscriptstyle \text{b}}$ = 35 au and a computation of 100 Myr.

One can clearly see that collisions between EPs and asteroids are greatly overestimated if we consider a merging approach. Indeed, we highlight that close to the MMR inside the HZ, the water transport to an EP's surface i.e. $\mathcal{W}_{\scriptscriptstyle \text{EP}}$ can be reduced significantly by almost 50\% for an Earth-, 68\% for a Mars- and 75\% for a Moon-sized object. We have the same statistics near the SR for a Mars- and Moon-sized except for an Earth as $\overline{v}_{\scriptscriptstyle \text{i}} = 3\,v_{\scriptscriptstyle \text{e}}$. In this case, $\mathcal{W}_{\scriptscriptstyle \text{EP}}$ is reduced by $\sim$ 30\%.  Even if no strong perturbation is located in the HZ (right panels), the real collision process shows that the incoming amount of water is reduced to $\mathcal{\widetilde{W}}_{\scriptscriptstyle \text{EP}}\,\sim 50\%\,\mathcal{W}_{\scriptscriptstyle \text{EP}}$ for a Moon-sized EP and to $\sim 20\%\,\mathcal{W}_{\scriptscriptstyle \text{EP}}$ for a Mars- and Earth-sized body (right panels).
A comparison of the left and respective right panels shows that around 1.0 au, $\mathcal{\widetilde{W}}_{\scriptscriptstyle \text{EP}}$ is nearly the same in both cases (SR $\in$ HZ and SR $\notin$ HZ) which indicates the importance of including SPH collisions into dynamical studies to avoid false effects as shown by the dotted lines of the left panels in Fig. \ref{F:wloss_collision}.

\begin{itemize}
\item [1.] The apparent positive aspect of orbital resonances is that due to eccentric motion of the EP at 1.0 au, the transported water ($\mathcal{W}_{\scriptscriptstyle \text{EP}}$) is boosted and ensures higher values than in case of circular motion of the EP. Moreover, if SR $\notin$ HZ, then even with a higher crossing frequency (see Fig. \ref{F:HZc}), the nearly circular motion of EPs close to 1.0 au limits the number of collisions with asteroids.

\item [2.] On the other hand, the negative aspect of high eccentric motion near 1.0 au is the high impact velocities which drastically reduces the efficiency of the water transport and leading to significantly  lower values of $\mathcal{\widetilde{W}}_{\scriptscriptstyle \text{EP}}$. It seems that nearly circular motion in the HZ is important to prevent relatively high water loss during collision as $\overline{v}_{\scriptscriptstyle \text{i}}$ are much lower when SR $\notin$ HZ.
 \end{itemize}

\section{Discussion: Other water loss processes}\label{S:discussion}

In this section, we discuss several processes that can significantly reduce the water content on the surface of an asteroid such as ice sublimation and atmospheric drag. 

\subsection{Ice sublimation}

Recently, the Rosetta mission revealed that comet 67P/CG sublimated 
$3.5\pm 0.5 \times  10^{28}$  H$_{2}$O molec/s after perihelion passage, implying a 
surface loss of 2--4 m \citep{hansen16}.
Water sublimation also occurs on the surface of asteroids, even in the 
main-belt, as shown by \cite{jewitt12}. We used the same approach as 
\cite{jewitt12} 
in order to estimate the surface recession rate of an asteroid due to ice sublimation by solving the following equation:

\begin{figure}[!h]
	\centering{
		\begin{tabular}{c}
			\includegraphics[angle=-90, width=\columnwidth]{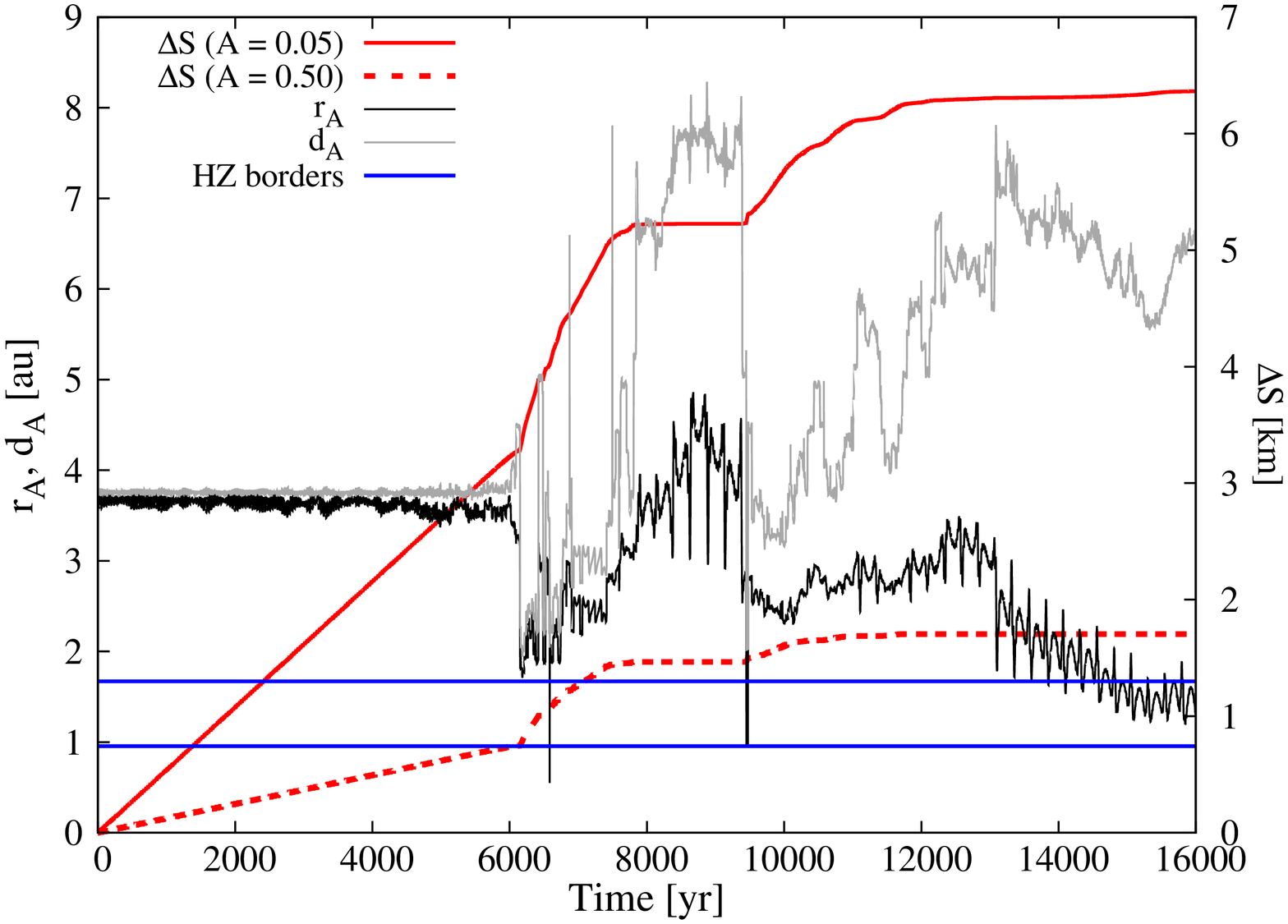} \\
			\includegraphics[angle=-90, width=\columnwidth]{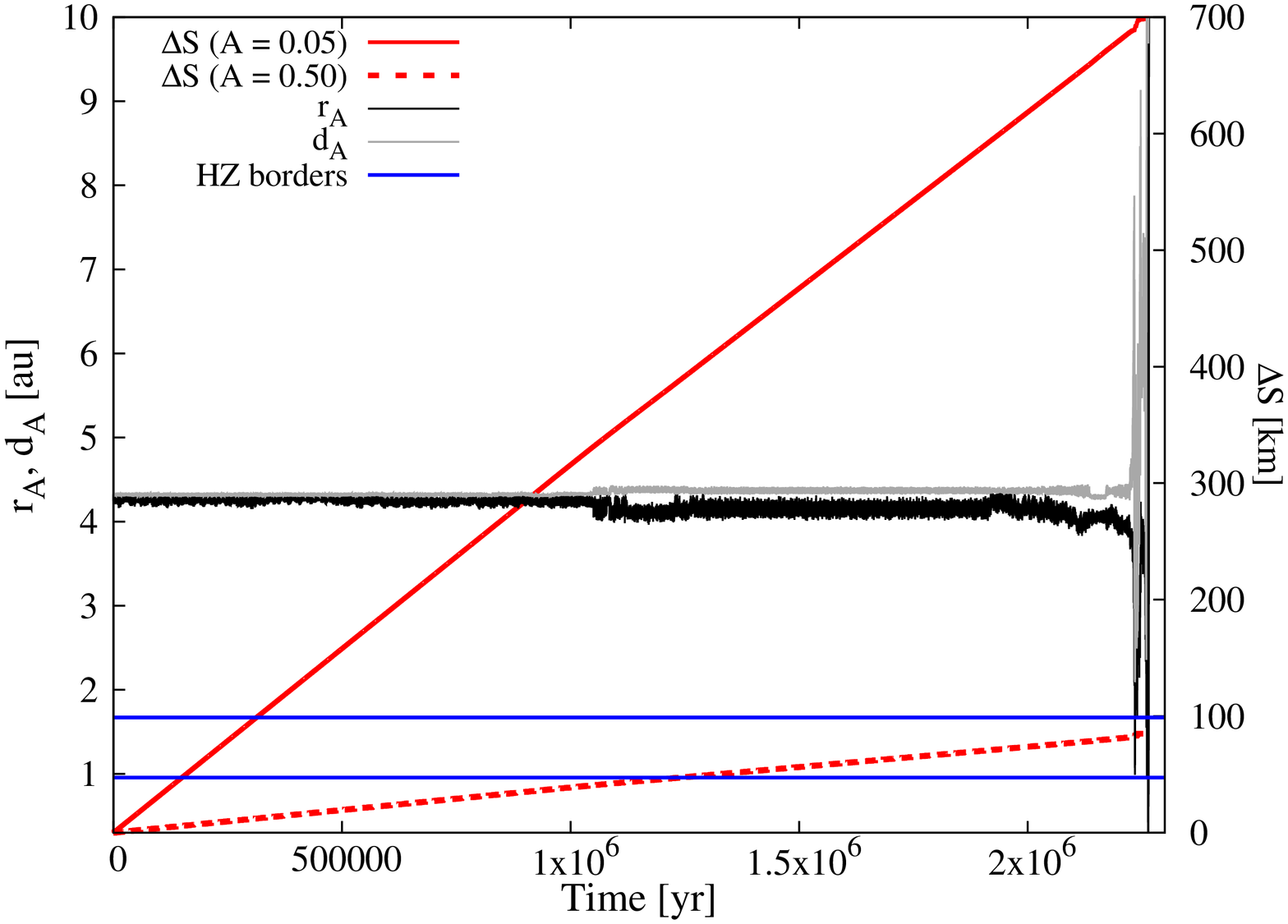}
	\end{tabular}}
	\caption{Evolution of the cumulative surface recession $\Delta$S for an albedo $A=0.05$ and $A=0.50$ (solid and dashed 
		lines, respectively). The heliocentric and mean distances $r_{\scriptscriptstyle {\text{A}}}$ and $d_{\scriptscriptstyle 
			{\text{A}}}$ are represented (black and grey lines, resp.) together with the HZ borders (blue 
		lines). Upper and bottom panels are for two asteroids with different orbital evolutions.}\label{F:dmdt_short}
\end{figure}


\begin{equation*}
\label{E:energy}
\displaystyle \frac{F_{\odot}(1-A)}{d^{\scriptscriptstyle 2}_{\scriptscriptstyle A} [\text{au}]}\cos{\theta} = 
\displaystyle \epsilon \sigma
T^{\scriptscriptstyle 4} + L(T)\,\rho_{\scriptscriptstyle {ice}}\,\frac{dS}{dt}(T)
\end{equation*}
\noindent
where
\begin{itemize}
	\item [1.] $F_{\odot}$ $\sim$ 1360 W.m$^{\scriptscriptstyle{-2}}$ is the solar constant;
	\item [2.] $A$, the bound albedo, is the product of the geometric albedo and the phase 
	integral and defines the fraction of the
	total incident stellar radiation reflected by an object back to space. 
	Asteroids can have \textit{A}$ \simeq$ 0.5,  but most of
	them have a relatively low albedo \citep{shestopalov11}. Ice material is known to be a good 
	radiation reflector but the sublimation process is more efficient when the ice is dirty ($A=0.05$) than clean 
	($A=0.50$). In our simulations, we considered both values for comparison;
	\item [3.] $d_{\scriptscriptstyle \text{A}} = \displaystyle \frac{q_{\scriptscriptstyle \text {A}} + Q_{\scriptscriptstyle \text {A}}}{2}$ is the mean distance to the primary star expressed in astronomical units and corresponds to the 
	average distance value between the asteroid's periapsis and apoapsis distance, $q_{\scriptscriptstyle \text {A}}$ and $Q_{\scriptscriptstyle \text {A}}$ respectively. We chose this definition in order to take 
	into account the non-circular motion of the asteroid. Indeed, using $r_{\scriptscriptstyle {\text {A}}}$ would 
	greatly overestimate the surface temperature of the asteroid in the case of an eccentric orbit, when integrating with time;
	\item [4.] $\theta$ is the angle between the Sun-direction and the normal of a patch of 
	the asteroid's surface. The surface temperature of the asteroid is minimized (assuming an isothermal surface) 
	when $\cos\theta = 1/4$ and maximized (reached at the sub-solar point on a non-rotating body) when $\cos\theta = 
	1$. In our simulation, we consider only the sub-solar case as the isothermal case 
	is negligible for main-belt asteroids;
	\item [5.] $\epsilon$ is the emissivity of the surface, $\epsilon \sim 0.9;$
	\item [6.] $\sigma = 5.67\times 10^{\scriptscriptstyle {-8}}$ W.m$^{-2}$.K$^{-4}$ is 
	the Boltzmann constant;
	\item [7.] $T$ is the equilibrium temperature at the surface expressed in K;
	\item [8.] $L$ is the latent heat of sublimation in J.kg$^{\scriptscriptstyle -1}$;
	\item [9.] $\rho_{\scriptscriptstyle {ice}}$ = 1000 kg.m$^{\scriptscriptstyle -3}$ is the ice density;
	\item [10.] $\displaystyle \frac{dS}{dt}$ is the surface recession rate (directly related to the surface mass loss rate) in 
	m.s$^{\scriptscriptstyle -1}$.
\end{itemize}
Our results are summarized in Fig. \ref{F:dmdt_short}, which shows the heliocentric 
and mean distances\footnote{these two quantities show how eccentric the orbit of 
the asteroid is when the solid gray and black lines do not overlap} $r_{\scriptscriptstyle {\text{A}}}$ and $d_{\scriptscriptstyle {\text{A}}}$ (black and grey lines, 
respectively), for two particular asteroids. In addition, we have plotted the cumulative surface recession $\displaystyle 
\Delta\,S=\sum\nolimits \frac{dS}{dt}\Delta\,t$, as a function of time, for $A = 0.05$ 
and $A=0.50$ (solid and dashed red lines, 
respectively). For the first asteroid (top panel), we can recognize several crosses with the HZ (solid blue lines), at $t\,\approx$  6700, 9700 and 14000 years, where collisions with an EP could occur. In case there is a collision at $\sim$ 1.0 au, this asteroid could lose between 1.5 and 6.5 km of its surface (according to the different values of the albedo $A$). The surface recession is moderate as the asteroid can collide no longer after the beginning of the integration. For the other asteroid (bottom panel), the first possible collision occurs 
at $t \approx 2.5\times 10^{\scriptscriptstyle 6}$ years where $\Delta$S is between 
90 and 700 km. In this case, it is very 
unlikely that this asteroid can deliver water initially located on its surface because its surface would become dry.  

\subsection{Atmospheric drag}
Planetary atmospheric properties depend strongly on the stage of formation of the 
EP. Indeed, during the disk-embedded phase of a protoplanet, primordial 
atmospheres are hot, dense structures which can extend to the protoplanet's Hill 
radius \citep{stokl16}. Because none of the terrestrial planets in the current state 
of our solar system show remnants of such primordial atmospheres, they probably must vanish at some time right after the disk phase. In order to 
estimate the mass loss $\Delta\,M$ of a Ceres-like object due to thermal 
ablation \citep{podolak88} caused by atmospheric drag, we proceed as 
follows.
\begin{itemize}
	\item [1.] we consider that the maximum mass loss accounts for a primordial 
	atmosphere extended to one Hill radius. As 
	suggested by \cite[in prep.]{ragossnig16}, we consider atmospheric parameters 
	(density and temperature) at 1 
	Myr in order to guarantee a quasi-stationary atmosphere. Furthermore, the typical 
	lifespan of the inner disk is believed to be around 1 to 3 Myr 
	\citep[see][]{williams11}, 
	which accounts for a regime where we assume a primordial atmosphere immediately 
	after the disk phase, with maximum temperature and density.  The body is then 
	dropped from one Hill radius with an initial 
	velocity such as the final impact velocity $\overline{v}_{\scriptscriptstyle \text{i}}$ of 
	the asteroid, after traveling through 
	the atmosphere, fits with our results presented in Sec. \ref{SS:vel}. The 
	variation of mass 
	$\Delta\,M$ evaluated 
	from one Hill radius to the EP's radius define the maximum mass loss. Due to the 
	low gravity of the Moon, we neglected 
	any effect of atmospheric pressure. We thus assume no mass loss in this case.\\
	
	\item [2.] Depending on the EP's mass, the evolution of the primordial atmosphere can 
	have different outcomes, as stated 
	earlier. We thus consider that the minimum mass loss accounts for (1) either a 
	present atmospheric model, as defined by 
	the International Standard Atmosphere (ISA) in the case of the Earth\footnote{in 
		this case, the atmospheric ISA values 
		were extrapolated up to one Hill radius} or (2) no presence of atmosphere around 	the EP, in the case of Mars- and the Moon-sized objects. In this latter case, we assume no mass loss.
\end{itemize}
\begin{figure}
	\centering{
		\includegraphics[angle=-90, width=\columnwidth]{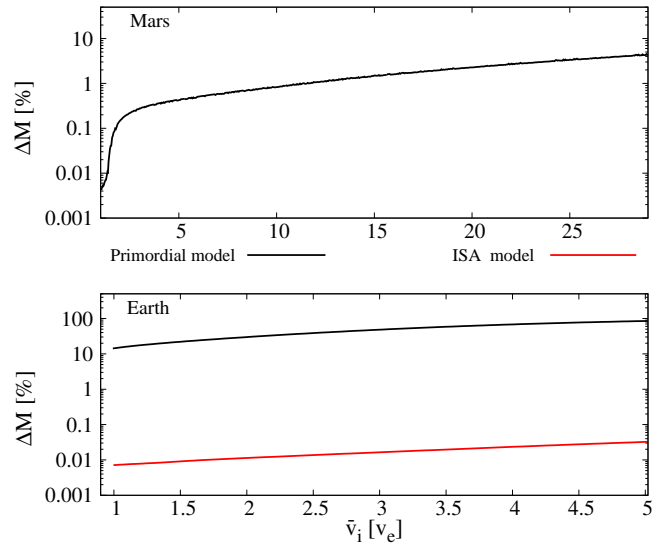}
	}
	\caption{Mass variation $\Delta\,M$ of a Ceres-like under a primordial 
		atmospheric pressure model (black line) for the 
		Earth and Mars, or an atmospheric model according to the ISA (red line), for the Earth only.}\label{F:atmos}
\end{figure}

Figure \ref{F:atmos} shows the mass loss $\Delta\,M$ of a Ceres-like body colliding with an Earth-type (bottom panel) or a 
Mars-type body (top panel) with different initial impact velocities $\overline{v}_{\scriptscriptstyle \text{i}}$. The black lines define the 
maximum mass loss due to the primordial atmosphere and the red line gives the minimum mass loss, for the Earth only, according to the ISA atmospheric model. As we can see, a Ceres-like asteroid entering the Hill radius of an Earth-like 
would lose a maximum of 15 -- 85\% of its initial mass, depending on its initial velocity. It is very unlikely that 
water can reach the EP's surface in this case\footnote{However, fragments can 
	remain in the EP's atmosphere}. A primordial atmosphere of a Mars-like would 
cause a maximum loss of 5\% 
for high impact velocities, allowing the asteroid to keep almost all of its initial water content. On a later stage of 
planetary formation, the atmospheric pressure according to the ISA parameters will have no effect on the asteroid's 
mass as the minimum mass loss is always below 0.1\%.

\section{Summary and conclusion}\label{S:conclusion}

In this paper, we investigated the influence of the location of orbital resonances (SR and MMRs) on the water transport 
onto EPs (Earth-, Mars- and Moon-sized objects) orbiting inside the habitable zone (HZ). For given binary configurations, the gas giant 
semi-major axis was chosen such that an SR lies either in the HZ at $\sim$ 1.0 au 
(causing high eccentric motion therein) or 
in the asteroid belt beyond the snow line at 2.7 au (and nearly circular motion in the HZ). We first showed that the frequency of 
asteroids reaching the HZ region (and even Mercury's distances) is much higher when the SR is inside the asteroid belt due to overlaps with MMRs. As a consequence, this produces a higher number of impactors on the EP's surface and therefore, a higher water transport. In addition, we showed that the presence of an SR and MMR inside the HZ could boost the water transport as the EPs can collide with more asteroids. This is apparently a positive mechanism for the water transport efficiency. However, our study shows clearly that dynamical results overestimate the water transport and need to be corrected by real simulations of collisions that provide the water loss due to an impact. Indeed, we showed that collisions in the HZ can occur with high impact velocities causing significant water loss for the asteroid (up to 100\%) when colliding at the EP's surface. The amount of water reaching the EP can be reduced by more than 50\%. 

During planetary formation, EPs also collide with each other and can transfer a fraction of their water content to each other. However, the merging approach generally used in simulations can overestimate the water content and mass of the surviving body resulting from the collision. SPH collision scenarios between embryo-sized objects have to be carried out in order to derive impact angles and velocity distributions, which will be the topic for a future work.

Even if water on the asteroid's surface may survive the collision, other mechanisms can significantly alter the wmf of asteroids, long before a collision. In this context, we were interested in ice sublimation, which can cause a severe recession of the asteroid's surface if they collide with an EP too long after the gas vanishes. Moreover, the primordial atmosphere of an EP can contribute up to 80\% of an asteroid's mass loss due to high impact velocity.

It is very likely that an asteroids' surface can become dry even before colliding in the HZ. Even if a water layer can resist, it might not survive the impact on the EP's surface. However, we cannot ignore that asteroids have indeed contributed to bring most of the water present on Earth's surface as 
the D/H ratio of chondrites asteroids matches with that of the Earth \citep{altwegg15}. Recently,  \cite{kuppers14} reported water emission from Ceres using $Herschel$ observations. However, there are uncertainties on the location of the ice layer even if \cite{formisano16} showed that an ice layer located only a few centimeters 
under the surface can explain the rate of produced water molecules. A possibility would be that the water of such icy objects is located beneath the surface and therefore, protected from early sublimation. This will be the subject of a future study.

\section*{Acknowledgements}
D.B, E.P.-L., T.I.M. and F.R. acknowledge the support of the Austrian Science Foundation 
(FWF) NFN project: Pathways to 
Habitability and related sub-projects S11608-N16 "Binary Star Systems and Habitability", S11603-N16 "Water transport" 
and S11604-N16 "Stars". D.B. and. E.P.-L. also acknowledge the Vienna Scientific Cluster (VSC project 70320) 
for computational resources.

\bibliographystyle{aas}

\bibliography{biblio} 



\end{document}